# The Time of Synchronization of Oscillations in Two Coupled Identical Subsystems

A. A. Koronovskiĭ*, A. E. Hramov**, and I. A. Khromova

*State Scientific Center "College", Saratov State University, Saratov, Russia*
*e-mail: * alkor@cas.ssu.runnet.ru; ** aeh@cas.ssu.runnet.ru*

**Abstract**—The mechanism of synchronization of oscillations in two identical coupled flow systems has been studied. The time (past the coupling onset) during which a synchronous oscillation regime is established depends on the oscillation phase difference between the subsystems. Variation of the coupling parameter leads to a change in the character of dependence of the synchronization time on the phase difference. Several types of synchronization processes are revealed, which are realized for various values of the coupling parameter.

In recent years, the problem of synchronization of nonlinear dynamical systems, in particular, the synchronization of chaotic oscillations, has received much attention. The original papers by Pecora and Carroll [1, 2] inspired extensive research in this direction and the number of publications still rapidly increases. This interest is related to the fact that the phenomenon of synchronization is involved in a large number of problems possessing both basic and applied significance. Examples include secret data transmission [3–6] and numerous problems in biology [7, 8], chemistry [9], ecology [10], astronomy [11], etc.

Besides the investigations aimed at determining the values of control parameters providing for the synchronization, establishing the conditions of synchronism breakage [12], and confirming weak coupling between several subsystems [13], the attention of researchers has been also drawn to determining the time of settling synchronous oscillations [14, 15]. The interest in this problem has both theoretical and practical aspects. For example, in the case of secret data transmission using chaotic synchronization techniques, the knowledge of the time interval during which the synchronization takes place allows the useful information to be correctly extracted [4]. It was demonstrated for two identical coupled van der Pol oscillators (representing the simplest model systems) [14, 15] that the dependence of the time of complete synchronization on the coupling parameter obeys a power law; the results were illustrated by numerical estimates.

This study was aimed at establishing how strongly (if at all) does the time of complete synchronization settling [6] in two identical coupled systems depends on the initial phase difference. Previously [17], we have demonstrated for two autooscillatory dynamical systems (van der Pol oscillator and distributed active medium of the "helical electron beam–backward electromagnetic wave" type) under the external periodic action that the time of synchronism settling at a frequency of the external driving signal significantly depends on the phase difference between natural oscillations and the driving signal.

Let us consider, following [14, 15], a system of two van der Pol oscillators with unidirectional coupling,

$$\ddot{x} + d(1-x^2)\dot{x} + x = 0,$$
$$\ddot{u} + d(1-u^2)\dot{u} + u = K(u-x)\delta(t-t_0), \quad (1)$$

where $x$ and $u$ are the dynamic variables characterizing the states of the first (driving) and second (driven) oscillator, respectively; $d = 0.3$ is the nonlinearity parameter; $K$ is the coupling parameter, $\delta(\xi)$ is the Heaviside function; and $t_0$ is the moment of switching on the unidirectional coupling, before which the two systems evolved independently (the time $t_0$ is selected sufficiently large for the transient processes in both systems to cease and the imaging points in the phase space to attain their limiting cycles).

The system of equations (1) was numerically solved using the fourth-order Runge–Kutta method with a time step of $h = 0.001$. The period of time $T_s$ to complete synchronization was determined as

$$T_s = t_s - t_0, \quad (2)$$

where $t_s$ is the moment of time corresponding to the onset of complete synchronism. A criterion for the complete synchronization was selected in the following form:

$$\sqrt{(x-u)^2 + (\dot{x}-\dot{u})^2} \leq 10^{-3}. \quad (3)$$

Figure 1 shows the characteristic plots of the period of time $T_s$, for which complete synchronism is settled in system (1) after switching on the coupling between two autooscillators, versus the initial phase $\varphi_x(t_0)$ of oscillations of the first (driving) oscillator. The moment of switching on the unidirectional coupling is chosen so that the phase $\varphi_u(t_0)$ of oscillations of the second (driven) oscillator would be zero. In fact, we study the dependence of the time to complete synchronization versus the initial phase difference $\varphi_x(t_0)-\varphi_u(t_0)$ between the driving and driven oscillations at the moment of coupling onset. The results of our investigations showed that a change in the initial phase $\varphi_u(t_0)$ of the driven oscillator leads to a shift of the $T_s$ versus $\varphi_x(t_0)$ curve along the phase axis in the absence of any qualitative changes in the character of this dependence.

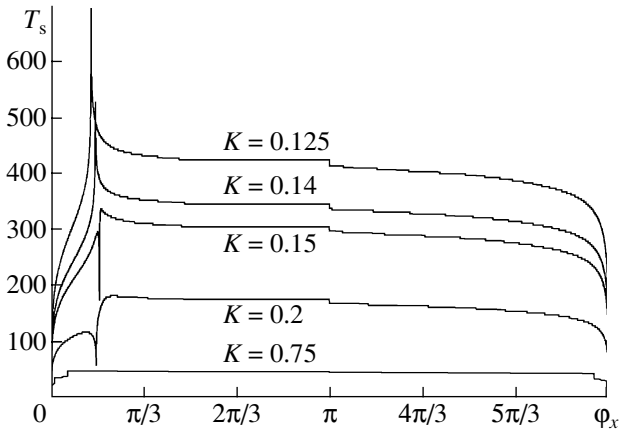

**Fig. 1.** Plots of the period of time $T_s$, for which complete synchronism is settled in system (1) after switching on the coupling between two autooscillators, versus the initial phase $\varphi_x(t_0)$ of oscillations of the first (driving) oscillator for various values of the coupling parameter $K$.

As can be seen from Fig. 1, the time to complete synchronization depends on the initial phase difference between driving and driven oscillators. Obviously, should the initial phase difference $\varphi_x(t_0)-\varphi_u(t_0)$ between the driving and driven oscillations be zero at the moment of coupling onset, the oscillations are synchronous from the very beginning and the time to synchronization is zero (Fig. 1).

For a nonzero initial phase difference, the time to complete synchronization of system (1) also differs from zero and changes rather weakly in a broad interval of variation of the initial phase $\varphi_x(t_0)$ of the driving oscillator. On the other hand, it can be clearly seen from the curves corresponding to small values of the coupling parameter $K$ that there are intervals of the initial phase $\varphi_x(t_0)$ in which the $T_s$ values significantly differ from the typical duration of synchronization. When the coupling parameters $K$ are below certain critical value $K_c$ (for the given control parameters, $K_c \approx 0.145$), there exists an initial phase of the driving oscillator $\varphi_x(t_0) = \varphi_c$ for which the time to synchronization significantly increases. When the coupling parameter $K$ increases above a certain critical level, the character of the $T_s$ versus $\varphi_x(t_0)$ curves exhibits a qualitative change (see Fig. 1, curves for $K = 0.14$ and 0.15), whereby approximately the same initial phase $\varphi_c$ of the driving oscillator brings a sharp minimum evidencing that the regime of complete synchronization at this point is attained significantly faster than in the typical case.[1] As the coupling parameter grows further, the time to complete synchronization decreases and all variations in the $T_s$ versus $\varphi_x(t_0)$ curves vanish (see Fig. 1, curve for $K = 0.75$).

Let us consider peculiarities of the complete synchronization settling in system (1) with $K = 0.15$. For this coupling parameter, the curve of the time to complete synchronization $T_s$ versus the initial phase of the driving oscillator $\varphi_x(t_0)$ exhibits a singularity at $\varphi_c = 0.547$ (Fig. 1). In the interval of initial phases of the driving oscillator $\varphi_x(t_0) \in (0, \varphi_c)$, the driven system behaves as follows (Fig. 2a). Initially, the oscillation amplitude sharply decreases and then gradually increases to approach the value corresponding to a stationary regime. This mechanism allows a mismatch between phases of the driving $\varphi_x(t_0)$ and driven $\varphi_u(t_0)$ subsystems to be eliminated. The phase trajectory reflecting the behavior of the driven subsystem winds on the limiting cycle from inside. A considerable fraction of time past the moment when the amplitude of oscillations in the driven system attained an initial value is spent for fine adjustment of the phase relation between the two oscillators.

For the initial phases of the driving oscillator $\varphi_x(t_0) \in (\varphi_c, 2\pi)$, the driven system behaves somewhat differently (Fig. 2c). It also sharply drops initially and then starts approaching the value corresponding to a stationary regime, but the phase trajectory winds on the limiting cycle from outside. The case of $\varphi_x(t_0) = \varphi_c$ corresponds to the boundary situation, whereby the system exhibits a change from the first to second scenario in behavior of the driven oscillator (Fog. 2b). This situation is optimum from the standpoint of synchronism settling: as soon as the amplitude of oscillations in the system reaches the value corresponding to a stationary regime, the phases of both subsystems turn out to be equal and no time is spent for their fine adjustment.

For smaller values of the coupling parameter $K$, the action of the first oscillator upon the second is insufficient to provide for realization of the above mechanism of the phase adjustment involving a sharp initial drop of the oscillation amplitude in the driven system, followed by gradual approach to the initial value and fine phase adjustment. With small $K$ values, the amplitude of oscillations in the driven subsystem changes only slightly and the phase difference between oscillations in the two subsystems is eliminated very slowly. This results in a general increase in the time $T_s$ required for synchronization and in a qualitative change in the char-acter of singularity (from minimum to maximum) at $\varphi_c$ in the $T_s$ versus $\varphi_x(t_0)$ curve.

---

[1] It should be noted that variation of the coupling parameter $K$ leads to small changes in the initial phase corresponding to a sharp difference of $T_s$ from the typical value.

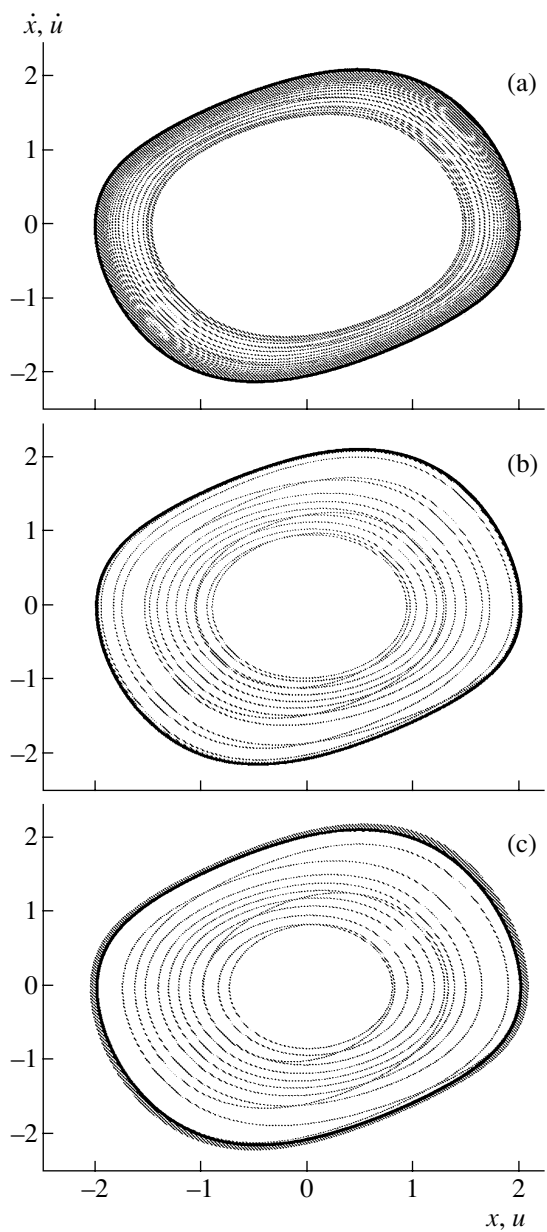

**Fig. 2.** Phase portraits illustrating the behavior of system (1) with the coupling parameter $K = 0.15$. The limiting cycle (thick solid curve) reflects behavior of the first (driving) oscillator in the coordinates $(x, \dot{x})$; light-gray phase trajectory shows the behavior of the second (driven) oscillator in the coordinates $(u, \dot{u})$ between the time moments $t_0$ and $t$. The initial phase of the driven oscillator is zero, while that of the driving oscillator is $\varphi_x(t_0) = 0.5$ (a), 0.547 (b), and 0.55 (c).

Finally, for greater values of the coupling parameter $K$ (see the curve for $K = 0.75$ in Fig. 1), the mechanism of synchronization settling is as follows. As the coupling is switched on, the driven subsystem exhibits a sharp change in the oscillation amplitude (initial drop followed by increase), at the expense of which the phase mismatch between the two subsystems is eliminated so that no fine adjustment is required. For this reason, the time to complete synchronization at large values of the coupling parameter $K$ is virtually independent of the initial phase difference between the driving and driven oscillators.

Thus, we have demonstrated that the time of complete synchronism settling in two identical subsystems depends on the initial phase difference between these subsystems. It was shown that there are several mechanisms of synchronization, which are realized for various values of the coupling parameter $K$.

**Acknowledgments.** This study was supported by the Federal Program "Integration" (project no. B0057), the Program of Support for Leading Scientific Schools in Russia, and the Scientific-Education Center "Nonlinear Dynamics and Biophysics" at the Saratov State University (Grant REC-006 from the U.S. Civilian Research and Development Foundation for the Independent States of the Former Soviet Union).